\documentclass[12pt]{article}

\usepackage[margin=1in]{geometry}
\usepackage{amsmath,amssymb,amsthm}
\usepackage{setspace}
\usepackage{hyperref}
\usepackage{booktabs}
\usepackage{longtable}
\usepackage{natbib}
\usepackage{graphicx}
\usepackage{microtype}
\usepackage{url}

\newtheorem{assumption}{Assumption}
\newtheorem{theorem}{Theorem}

\title{Correcting Sample Selection Bias in PISA Rankings}
\author{Onil Boussim\\
  \small Department of Economics, The Pennsylvania State University, University Park, USA\\
  \small \href{}{oib5044@psu.edu}}
\date{}

\begin{document}

\maketitle

\begin{abstract}
This paper proposes a method to account for sample selection (survivor) bias in cross-country
comparisons. International assessments such as the Programme for International Student Assessment
(PISA) observe outcomes only for students enrolled in school at age 15, which can distort
comparisons in countries with high dropout rates. I consider a quantile-based selection correction
that delivers bounds on countries' average performance rather than point estimates. Using these
bounds, I construct optimistic and pessimistic rankings, with each country's true rank lying between
the two. An application to PISA 2018 shows that correcting for selection bias leads to substantial
changes in countries' average scores and rankings.
\end{abstract}



%% ----------------------------------------------------------------
\section{Introduction}
%% ----------------------------------------------------------------

International comparisons of educational achievement, typically based on standardized assessments,
are essential for evaluating and shaping education policies worldwide. Such comparisons allow
policymakers to benchmark national education systems, assess reforms, and identify best practices,
with rankings often influencing funding, curriculum changes, and broader strategic decisions
\citep{Nagy1996,Martin2000,McEwan2004,Cromley2009,Tienken2008,McGaw2008,Jakubowski2015}.
Among these assessments, the Programme for International Student Assessment (PISA) is the most
widely recognized, evaluating 15-year-old students every three years in reading, mathematics, and
science. PISA emphasizes the application of knowledge to real-world problems rather than mastery of
a prescribed curriculum, and participation includes approximately 80 countries, each selecting a
nationally representative sample of 4,500 to 10,000 students across 150 to 250 schools to capture
socioeconomic diversity.

PISA results have driven significant educational reforms. Germany's low ranking in 2001
prompted nationwide policy changes that improved outcomes by 2012
\citep{Knodel2013,Ringarp2016}, while countries such as Sweden, Canada, and Norway have leveraged
PISA to guide investments and evaluate reforms \citep{Ringarp2016,Knighton2010,Stray2020}.

Despite its prominence, PISA assessments suffer from a fundamental sample selection problem: not all
students in the target age group are observed. Students may drop out or be excluded for logistical
reasons, creating potential bias if the unobserved population differs systematically from
participants. For instance, countries that disproportionately exclude low-performing students may
appear to outperform some countries with near-complete coverage, and the reverse may occur if
high-performing students are missing. The magnitude and direction of this bias depend on the
characteristics of the excluded groups, suggesting that conventional rankings may reflect differences
in sample composition rather than true educational performance
\citep{Rotberg1995,Berliner1993,Ferreira2014}.

Solving this selection problem raises two key econometric challenges. First, there is no information
on excluded individuals, unlike typical labor market settings where surveys observe covariates for
both participants and non-participants \citep{Heckman1974,Arellano2017,Chernozhukov2025}. Second,
most nonparametric approaches to selection correction rely on valid instruments, which are not
available in the PISA context. Together, these features severely limit identification. To address
this, I adopt a partial identification approach that does not aim to recover the exact average
performance of all students, but instead derives upper and lower bounds using observed outcomes and
known coverage rates.

This paper contributes to the literature on partial identification under sample selection.
\citet{Lee2009} derives sharp nonparametric bounds on treatment effects under monotonicity using a
trimming approach. In contrast, our setting lacks treatment variation and leverages coverage rates
reported by PISA rather than focusing on always-observed units. \citet{Honore2020,Honore2024} study
selection models without exclusion restrictions under parametric or semiparametric assumptions,
whereas our approach is fully nonparametric and relies only on coverage rates and a stochastic
dominance assumption. \citet{Fan2018} develop bounds for censored quantile regression using
parametric copulas, while we impose no parametric structure. \citet{Fan2009,Fan2012} apply
Fr\'{e}chet--Hoeffding bounds to quantile treatment effects; although we use similar tools, we bound
quantiles of a latent outcome rather than treatment effects. Finally, the paper relates to the
literature on non-additive quantile selection models. \citet{Arellano2017} derive related bounds
using continuous instruments, whereas our framework does not rely on instruments.

The upper bound represents a best-case scenario \emph{relative to Assumption~\ref{ass:dominance}},
assuming that excluded students perform just like those who took the test. Note that an even more
optimistic scenario---where excluded students outperform those tested---is ruled out by this
Assumption, which we regard as empirically plausible. The lower bound
represents a worst-case scenario, assuming that excluded students are systematically less skilled
than participants. Together, these bounds define a range of plausible values for a country's average
performance. This approach allows policymakers to see how much uncertainty selection introduces into
rankings.

Applying this methodology to PISA 2018 demonstrates that correcting for sample selection affects
international rankings: countries with low coverage often see downward adjustments, while those with
near-complete coverage remain largely unaffected. The remainder of the paper is organized as
follows. Section~\ref{sec:model} presents the econometric model and identification strategy.
Section~\ref{sec:application} applies the method to PISA 2018 data and discusses the findings.
Section~\ref{sec:conclusion} concludes.

%% ----------------------------------------------------------------
\section{Model and Identification}\label{sec:model}
%% ----------------------------------------------------------------

\subsection{Analytical Framework}

Consider a country and a single cohort of students targeted by PISA, which assesses students at
age 15. A major challenge in interpreting the test results is that not all individuals in the target
age group are observed; only those still enrolled in school at the time of the assessment are
included. This introduces survivor, or sample selection, bias, which can be especially pronounced in
countries with high dropout rates.

To formalize this issue, let $Y^*$ denote the potential assessment score an individual would obtain if assessed,
regardless of whether they are observed. The observed assessment score,
denoted by $Y$, is available only for individuals who meet the inclusion criteria. Since assessment
scores are continuous, the distribution of $Y^*$ can be represented using a quantile (or rank)
formulation. Let $U \sim \mathcal{U}[0,1]$ denote the individual's rank in the distribution of
potential scores. Because the outcome is continuous, we can write: $Y^* = Q_{Y^*}(U)$, where
$U = F_{Y^*}(Y^*) \sim \mathcal{U}[0,1]$. Next, define a binary selection indicator $S$, taking value 1 if the individual is
included in the assessment (i.e., enrolled at age 15) and 0 otherwise. When $S=1$, the
researcher observes the assessment score; when $S=0$, the score is completely unobserved. Formally,
$Y = Y^*$ if $S=1$, and $Y$ is undefined when $S=0$. From a policy perspective, the relevant object
of interest is the mean of $Y^*$, which captures overall educational performance net of selection.

\subsection{Theoretical Results}\label{sec:theory}

In practice, the quantile function $Q_{Y\mid S=1}(\cdot)$ can be directly estimated from the
observed data. However, because we have no information on the outcomes of excluded individuals,
identification depends on limited but essential auxiliary information. The first key assumption
relates to the coverage rate.

\begin{assumption}[Identification of coverage rate]\label{ass:coverage}
The probability of inclusion $p \equiv \mathbb{P}(S=1)$ is identified.
\end{assumption}

This assumption is mild and typically satisfied in practice. In the case of PISA, for example,
official documentation reports the proportion of the age-eligible population that is enrolled in
school and thus eligible to be assessed. Using basic probability restrictions, we can derive
informative bounds. For $u \in [0,1]$, define:
\[
  \tilde{u} \equiv \mathbb{P}(U \leq u \mid S=1) = \frac{\mathbb{P}(U \leq u,\, S=1)}{p}.
\]
This quantity captures how ranks in the full population map into ranks among observed students. The
first key result establishes a simple relationship between the quantiles of potential and observed
outcomes. The intuition is straightforward: the observed score distribution corresponds to a
truncated version of the full distribution, consisting only of individuals with $S=1$. Once we know
how a given population rank $u$ translates into a rank $\tilde{u}$ among observed students, we can
recover the corresponding quantile of $Y^*$ from the observed data. To see why, fix $u \in [0,1]$
and consider:
\[
  \tilde{u}
  = \mathbb{P}(U \leq u \mid S=1)
  = \mathbb{P}(Y^* \leq Q_{Y^*}(u) \mid S=1)
  = F_{Y\mid S=1}\!\bigl(Q_{Y^*}(u)\bigr).
\]
Applying the observed quantile function yields $Q_{Y^*}(u) = Q_{Y\mid S=1}(\tilde{u})$.

Focusing on the corrected rank $\tilde{u}$ and applying the Fr\'{e}chet--Hoeffding bounds to the
joint probability gives:
\[
  \frac{\max\{u + p - 1,\, 0\}}{p} \;\leq\; \tilde{u} \;\leq\; \frac{\min\{u,\, p\}}{p}.
\]
Because $Q_{Y\mid S=1}(\cdot)$ is monotone, these bounds translate directly into bounds on the
potential outcome quantiles:
\[
  Q_{Y\mid S=1}\!\!\left(\frac{\max\{u+p-1,0\}}{p}\right)
  \;\leq\; Q_{Y^*}(u)
  \;\leq\; Q_{Y\mid S=1}\!\!\left(\frac{\min\{u,p\}}{p}\right).
\]
These bounds are fully nonparametric and require no behavioral assumptions. However, they can be
wide and therefore uninformative, especially when coverage rates are low. To tighten them, I
introduce the following weak and policy-relevant assumption:

\begin{assumption}[Stochastic Dominance]\label{ass:dominance}
For all $u \in [0,1]$, $Q_{Y^*\mid S=0}(u) \leq Q_{Y^*\mid S=1}(u)$.
\end{assumption}

This assumption states that the distribution of potential scores for observed students first-order
stochastically dominates that of excluded students. In intuitive terms, individuals who remain
enrolled until the assessment age tend to have higher latent academic ability than those who drop
out earlier. This assumption is consistent with a wide range of educational and socioeconomic
mechanisms documented in the education policy literature. Under this assumption, the upper bound
simplifies substantially.

\begin{assumption}[Bounded support from below]\label{ass:support}
The infimum of the support of $Y\mid S=1$, denoted $Q_{Y\mid S=1}(0)$, is finite.
\end{assumption}

In the PISA context, this assumption is natural: PISA scores lie on a fixed scale and are bounded
below. We use the minimum PISA score for each country as an estimate, because it is a consistent
estimator for the infimum of the support and the sample sizes are large enough.

\begin{theorem}[Partial Identification]\label{thm:main}
Under Assumptions~\ref{ass:coverage}, ~\ref{ass:dominance} and ~\ref{ass:support} , the following bounds are valid and
sharp for all $u \in [0,1]$:
\[
  Q_{Y\mid S=1}\!\!\left(\frac{\max\{u+p-1,0\}}{p}\right)
  \;\leq\; Q_{Y^*}(u)
  \;\leq\; Q_{Y\mid S=1}(u).
\]
Moreover, the population mean satisfies:
\[
  (1-p)\,Q_{Y\mid S=1}(0) + p\,\mathbb{E}(Y\mid S=1)
  \;\leq\; \mathbb{E}(Y^*)
  \;\leq\; \mathbb{E}(Y\mid S=1).
\]
\end{theorem}

The proof is provided in the Appendix. The partial identification framework provides an intuitive
way to account for survivor bias. The upper bound $\mathbb{E}(Y\mid S=1)$ corresponds to the
optimistic benchmark---the scenario in which excluded individuals would perform just as well as
observed students. This considered optimistic because enrolled students are more likely to be perform better than non-enrolled. The lower bound represents the pessimistic scenario in which unobserved students
perform worse. The true population mean $\mathbb{E}(Y^*)$ must lie between these two values.

%% ----------------------------------------------------------------
\section{Application}\label{sec:application}
%% ----------------------------------------------------------------

In this section, I apply the correction and derive rankings using the PISA 2018 assessment data.
The analysis is based on a sample of 77 countries and the full table is available in the appendix.
The variable $p$ denotes the coverage rate. \textit{Lbound} and \textit{Ubound} provide the lower
and upper bound means. \textit{Rank~1} refers to the official ranking based on reported PISA 2018
mean scores (coinciding with the ranking by upper bound). \textit{Rank~2} represents the ranking
based on the lower bound. The width of the bounds on $\mathbb{E}(Y^*)$ is given by:
\[
  \text{Width}
  = \mathbb{E}(Y\mid S=1) - \text{lower bound}
  = (1-p)\bigl(\mathbb{E}(Y\mid S=1) - Q_{Y\mid S=1}(0)\bigr).
\]
This expression is increasing in $(1-p)$: as $p$ approaches 1, the bounds collapse to a point and
$\mathbb{E}(Y^*)$ is point-identified. Conversely, for countries with low coverage rates, the bounds
are wide. Approximately 40\% of countries in our sample have $p \geq 0.90$, for which the bounds are
tight. Countries with $p < 0.70$ (such as Panama, Philippines, and Jordan) have substantially wider
bounds, reflecting greater uncertainty about the performance of excluded students.

Rank~2 is obtained by ranking all countries simultaneously by their lower bound means. This provides
a coherent, symmetric comparison across all countries---a simultaneous pessimistic scenario.
However, we note that from the perspective of a specific country $j$, the worst-case scenario is
more extreme: it corresponds to $j$ being ranked using its lower bound while all competitor
countries are ranked using their upper bounds. The simultaneous lower-bound ranking is more
informative for systemic policy comparisons, while the individual worst-case ranks are more
conservative.

The PISA 2018 mathematics rankings reveal clear patterns across countries. East Asian countries such
as China, Singapore, Hong Kong, Taiwan, Japan, and Korea consistently occupy the top positions.
China holds Rank~1\,=\,1 and Rank~2\,=\,3, indicating high performance with moderate sensitivity to
selection correction. Germany is notable with Rank~1\,=\,20 and Rank~2\,=\,6, indicating that
Germany's near-complete coverage ($p = 0.993$) means its lower bound is very close to its observed
mean, so it moves up sharply in the corrected ranking. In reading, East Asian countries remain top
performers; European countries such as Finland, Ireland, and Estonia also perform well with small
rank spreads.

%% ----------------------------------------------------------------
\section{Conclusion}\label{sec:conclusion}
%% ----------------------------------------------------------------

In this paper, I propose a method to address sample selection bias in cross-country comparisons
using data from international assessments such as PISA. The proposed correction leverages a quantile
selection model, which, under natural assumptions, enables partial identification of latent quantiles
and, consequently, the latent means. The application of this method to the PISA 2018 data reveals
that the rankings can change.

%% ----------------------------------------------------------------
\section*{Acknowledgements}
%% ----------------------------------------------------------------

I am grateful to Marc Henry, Andres Aradillas-Lopez, Michael Gechter, Ismael Mouriﬁ\'{e}, and all
participants of the African Econometric Society 2024 for valuable suggestions and comments. All
errors are mine.

%% ----------------------------------------------------------------
\section*{Conflict of Interest Statement}
%% ----------------------------------------------------------------

The author has no conflict of interest to declare.

%% ----------------------------------------------------------------
\section*{Data Availability Statement}
%% ----------------------------------------------------------------

The data that support the empirical findings of this study are openly available at:
\url{https://www.oecd.org/en/data/datasets/pisa-2018-database.html}

%% -----------------------------------------------------------

%% ----------------------------------------------------------------
\bibliographystyle{aer}

%% ----------------------------------------------------------------
\appendix
%% ----------------------------------------------------------------

\section*{Appendix}

\subsection*{Table 1: PISA 2018 Mathematics}

\begin{longtable}{lrrrrr}
\toprule
Country & $p$ & Lbound & Ubound & Rank 1 & Rank 2 \\
\midrule
\endfirsthead
\toprule
Country & $p$ & Lbound & Ubound & Rank 1 & Rank 2 \\
\midrule
\endhead
\midrule \multicolumn{6}{r}{\textit{Continued on next page}} \\
\endfoot
\bottomrule
\endlastfoot
China               & 0.812 & 524.98 & 591.13 & 1  & 3  \\
Singapore           & 0.953 & 550.87 & 566.76 & 2  & 1  \\
Macau               & 0.883 & 512.48 & 556.64 & 3  & 4  \\
Hong Kong           & 0.984 & 547.15 & 551.78 & 4  & 2  \\
Taiwan              & 0.921 & 501.30 & 531.97 & 5  & 7  \\
Japan               & 0.909 & 501.89 & 528.62 & 6  & 6  \\
Korea               & 0.881 & 482.78 & 525.15 & 7  & 9  \\
Estonia             & 0.931 & 502.42 & 522.99 & 8  & 5  \\
Netherlands         & 0.912 & 490.21 & 520.53 & 9  & 8  \\
Poland              & 0.900 & 486.89 & 516.44 & 10 & 10 \\
Switzerland         & 0.889 & 475.49 & 515.39 & 11 & 11 \\
Canada              & 0.863 & 468.60 & 512.03 & 12 & 15 \\
Denmark             & 0.878 & 477.75 & 510.21 & 13 & 12 \\
Slovenia            & 0.979 & 501.00 & 508.90 & 14 & 7  \\
Belgium             & 0.936 & 488.35 & 508.28 & 15 & 9  \\
Finland             & 0.963 & 497.06 & 507.84 & 16 & 8  \\
Norway              & 0.911 & 471.94 & 502.64 & 17 & 13 \\
Sweden              & 0.857 & 460.94 & 502.55 & 18 & 18 \\
United Kingdom      & 0.848 & 444.36 & 502.20 & 19 & 23 \\
Germany             & 0.993 & 498.61 & 499.96 & 20 & 6  \\
Austria             & 0.889 & 465.62 & 499.47 & 21 & 16 \\
Ireland             & 0.962 & 489.24 & 499.25 & 22 & 10 \\
Czech Republic      & 0.954 & 485.91 & 498.94 & 23 & 11 \\
Latvia              & 0.886 & 469.22 & 497.19 & 24 & 17 \\
France              & 0.913 & 470.20 & 495.58 & 25 & 14 \\
Iceland             & 0.916 & 471.74 & 495.07 & 26 & 13 \\
New Zealand         & 0.888 & 463.03 & 494.61 & 27 & 19 \\
Portugal            & 0.873 & 450.56 & 492.99 & 28 & 22 \\
Australia           & 0.894 & 457.54 & 492.15 & 29 & 20 \\
Russia              & 0.936 & 469.73 & 487.92 & 30 & 12 \\
Slovak Republic     & 0.862 & 436.41 & 487.60 & 31 & 28 \\
Italy               & 0.846 & 438.56 & 486.28 & 32 & 27 \\
Lithuania           & 0.903 & 456.12 & 484.15 & 33 & 21 \\
Luxembourg          & 0.871 & 441.17 & 483.50 & 34 & 25 \\
Hungary             & 0.896 & 444.36 & 482.26 & 35 & 23 \\
United States       & 0.861 & 435.59 & 477.92 & 37 & 30 \\
Belarus             & 0.876 & 431.22 & 471.94 & 38 & 33 \\
Malta               & 0.972 & 461.70 & 470.86 & 39 & 24 \\
Croatia             & 0.891 & 433.32 & 464.43 & 40 & 32 \\
Israel              & 0.809 & 395.24 & 462.21 & 41 & 38 \\
Turkey              & 0.726 & 372.29 & 453.63 & 42 & 41 \\
Ukraine             & 0.867 & 409.65 & 453.17 & 43 & 37 \\
Greece              & 0.927 & 429.60 & 450.96 & 44 & 35 \\
Serbia              & 0.885 & 413.42 & 448.25 & 45 & 39 \\
Malaysia            & 0.723 & 361.52 & 440.57 & 46 & 44 \\
United Arab Emirates & 0.918 & 413.32 & 437.08 & 48 & 39 \\
Albania             & 0.757 & 363.13 & 436.71 & 49 & 43 \\
Romania             & 0.726 & 339.93 & 430.68 & 50 & 50 \\
Bosnia-Herzegovina  & 0.823 & 362.94 & 406.85 & 61 & 46 \\
Mexico              & 0.664 & 294.50 & 408.64 & 60 & 52 \\
Georgia             & 0.826 & 349.84 & 398.70 & 65 & 48 \\
Peru                & 0.731 & 324.00 & 399.30 & 64 & 51 \\
North Macedonia     & 0.947 & 377.12 & 393.20 & 67 & 45 \\
Colombia            & 0.619 & 298.78 & 391.13 & 68 & 53 \\
Brazil              & 0.650 & 297.08 & 382.82 & 69 & 54 \\
Argentina           & 0.806 & 331.35 & 379.69 & 70 & 47 \\
Indonesia           & 0.849 & 335.57 & 378.05 & 71 & 49 \\
Saudi Arabia        & 0.845 & 336.42 & 374.13 & 72 & 50 \\
Morocco             & 0.643 & 284.42 & 369.02 & 73 & 55 \\
Kosovo              & 0.844 & 322.43 & 364.91 & 74 & 51 \\
Panama              & 0.535 & 242.69 & 352.43 & 75 & 56 \\
Philippines         & 0.679 & 267.10 & 352.39 & 76 & 57 \\
Dominican Republic  & 0.730 & 261.21 & 324.53 & 77 & 58 \\
\end{longtable}

\subsection*{Table 2: PISA 2018 Reading}

\begin{longtable}{lrrrrr}
\toprule
Country & $p$ & Lbound & Ubound & Rank 1 & Rank 2 \\
\midrule
\endfirsthead
\toprule
Country & $p$ & Lbound & Ubound & Rank 1 & Rank 2 \\
\midrule
\endhead
\midrule \multicolumn{6}{r}{\textit{Continued on next page}} \\
\endfoot
\bottomrule
\endlastfoot
China               & 0.812 & 493.82 & 555.25 & 1  & 3  \\
Singapore           & 0.953 & 532.06 & 549.50 & 2  & 1  \\
Macau               & 0.883 & 484.20 & 525.13 & 3  & 4  \\
Hong Kong           & 0.984 & 519.11 & 524.32 & 4  & 2  \\
Estonia             & 0.931 & 503.73 & 523.37 & 5  & 5  \\
Canada              & 0.863 & 475.63 & 520.12 & 6  & 8  \\
Finland             & 0.963 & 506.40 & 519.66 & 7  & 6  \\
Ireland             & 0.962 & 507.04 & 518.41 & 8  & 7  \\
Korea               & 0.881 & 475.92 & 513.84 & 9  & 9  \\
Poland              & 0.900 & 478.76 & 512.16 & 10 & 9  \\
Sweden              & 0.857 & 453.82 & 505.74 & 11 & 13 \\
New Zealand         & 0.888 & 470.81 & 505.29 & 12 & 10 \\
United States       & 0.861 & 459.64 & 505.00 & 13 & 15 \\
United Kingdom      & 0.848 & 460.17 & 504.34 & 14 & 14 \\
Japan               & 0.909 & 471.65 & 503.30 & 15 & 11 \\
Taiwan              & 0.921 & 475.98 & 502.37 & 16 & 11 \\
Australia           & 0.894 & 465.19 & 502.27 & 17 & 13 \\
Denmark             & 0.878 & 464.52 & 501.88 & 18 & 13 \\
Norway              & 0.911 & 468.84 & 499.11 & 19 & 13 \\
Germany             & 0.993 & 497.36 & 498.94 & 20 & 8  \\
Slovenia            & 0.979 & 490.11 & 495.64 & 21 & 9  \\
France              & 0.913 & 464.55 & 493.03 & 22 & 13 \\
Belgium             & 0.936 & 472.32 & 492.87 & 23 & 13 \\
Portugal            & 0.873 & 451.54 & 491.95 & 24 & 15 \\
Czech Republic      & 0.954 & 477.09 & 490.42 & 25 & 10 \\
Netherlands         & 0.912 & 454.31 & 484.58 & 26 & 15 \\
Austria             & 0.889 & 448.66 & 483.53 & 27 & 15 \\
Switzerland         & 0.889 & 448.99 & 483.50 & 28 & 15 \\
Croatia             & 0.891 & 446.23 & 479.08 & 29 & 16 \\
Russia              & 0.936 & 460.20 & 478.71 & 30 & 15 \\
Latvia              & 0.886 & 447.13 & 478.05 & 31 & 16 \\
Spain               & 0.918 & 452.01 & 476.58 & 32 & 15 \\
Italy               & 0.846 & 426.62 & 476.17 & 33 & 18 \\
Hungary             & 0.896 & 445.16 & 476.10 & 34 & 17 \\
Lithuania           & 0.903 & 448.38 & 475.75 & 35 & 16 \\
Belarus             & 0.876 & 438.37 & 474.12 & 36 & 17 \\
Iceland             & 0.916 & 448.49 & 473.80 & 37 & 16 \\
Israel              & 0.809 & 404.81 & 469.99 & 38 & 21 \\
Luxembourg          & 0.871 & 430.47 & 469.69 & 39 & 19 \\
Ukraine             & 0.867 & 422.09 & 465.42 & 40 & 20 \\
Turkey              & 0.726 & 387.50 & 465.23 & 41 & 21 \\
Slovak Republic     & 0.862 & 419.95 & 457.78 & 42 & 20 \\
Greece              & 0.927 & 436.51 & 457.30 & 43 & 20 \\
Chile               & 0.893 & 420.92 & 452.76 & 44 & 20 \\
Malta               & 0.972 & 439.68 & 448.28 & 45 & 20 \\
Serbia              & 0.885 & 409.20 & 439.25 & 46 & 21 \\
United Arab Emirates & 0.918 & 407.66 & 431.05 & 47 & 21 \\
Costa Rica          & 0.628 & 337.61 & 427.14 & 48 & 23 \\
Romania             & 0.726 & 323.94 & 427.05 & 49 & 23 \\
Uruguay             & 0.780 & 362.02 & 426.45 & 50 & 22 \\
Moldova             & 0.951 & 409.71 & 423.94 & 51 & 21 \\
Montenegro          & 0.947 & 407.86 & 420.95 & 52 & 21 \\
Mexico              & 0.664 & 339.31 & 420.60 & 53 & 23 \\
Bulgaria            & 0.720 & 348.51 & 419.90 & 54 & 23 \\
Jordan              & 0.540 & 262.10 & 419.01 & 55 & 25 \\
Malaysia            & 0.723 & 348.04 & 415.01 & 56 & 23 \\
Brazil              & 0.650 & 312.14 & 413.13 & 57 & 24 \\
Colombia            & 0.619 & 313.85 & 412.12 & 58 & 24 \\
Brunei              & 0.974 & 403.21 & 409.04 & 59 & 21 \\
Qatar               & 0.923 & 386.21 & 407.09 & 60 & 22 \\
Albania             & 0.757 & 344.93 & 405.35 & 61 & 23 \\
Bosnia-Herzegovina  & 0.823 & 361.45 & 402.91 & 62 & 22 \\
Argentina           & 0.806 & 338.80 & 401.19 & 63 & 24 \\
Peru                & 0.731 & 331.68 & 400.32 & 64 & 24 \\
Saudi Arabia        & 0.845 & 356.32 & 398.93 & 65 & 23 \\
Thailand            & 0.724 & 327.11 & 393.27 & 66 & 24 \\
North Macedonia     & 0.947 & 374.82 & 392.09 & 67 & 22 \\
Azerbaijan          & 0.463 & 264.30 & 389.46 & 68 & 25 \\
Kazakhstan          & 0.920 & 369.82 & 386.67 & 69 & 22 \\
Georgia             & 0.826 & 338.46 & 379.69 & 70 & 24 \\
Panama              & 0.535 & 259.74 & 378.23 & 71 & 25 \\
Indonesia           & 0.849 & 341.16 & 370.96 & 72 & 24 \\
Morocco             & 0.643 & 286.39 & 359.63 & 73 & 25 \\
Lebanon             & 0.867 & 306.62 & 352.80 & 74 & 25 \\
Kosovo              & 0.844 & 321.98 & 352.50 & 75 & 25 \\
Dominican Republic  & 0.730 & 279.38 & 341.09 & 76 & 25 \\
Philippines         & 0.679 & 282.68 & 339.47 & 77 & 25 \\
\end{longtable}

\subsection*{Proofs of the Results in the Main Text}

\textbf{Step 1: Validity of the inequalities.}
By the Fr\'{e}chet inequality, we already know that:
\[
  Q_{Y^*}(u) \geq Q_{Y\mid S=1}\!\!\left(\frac{\max\{u+p-1,0\}}{p}\right).
\]
Now consider:
\[
  u = \mathbb{P}(U \leq u \mid S=1)\,p + \mathbb{P}(U \leq u \mid S=0)(1-p).
\]
Assumption~\ref{ass:dominance} (stochastic dominance) is equivalent to:
\[
  \mathbb{P}(U \leq u \mid S=1) \leq \mathbb{P}(U \leq u \mid S=0).
\]
Knowing that:
\[
  \mathbb{P}(U \leq u \mid S=1)\,p + \mathbb{P}(U \leq u \mid S=1)(1-p)
  = \mathbb{P}(U \leq u \mid S=1) = \tilde{u} \leq u,
\]
and using the monotonicity of $Q_{Y\mid S=1}$, we have:
\[
  Q_{Y\mid S=1}(\tilde{u}) \leq Q_{Y\mid S=1}(u)
  \quad\Longleftrightarrow\quad
  Q_{Y^*}(u) \leq Q_{Y\mid S=1}(u).
\]

\textbf{Step 2: Sharpness.}
For the lower bound, consider $S = \mathbf{1}\{U \geq 1-p\}$. Then $\mathbb{P}(U \geq 1-p) = p$
and:
\[
  \tilde{u} = \frac{\mathbb{P}(U \leq u,\, U \geq 1-p)}{p}
  = \frac{\max\{u+p-1,0\}}{p}.
\]
Stochastic dominance also holds:
\[
  \frac{\max\{u+p-1,0\}}{p}
  \leq \mathbb{P}(U \leq u \mid U < 1-p)
  = \frac{\min\{u,1-p\}}{1-p},
\]
which implies $\mathbb{P}(U \leq u \mid S=1) \leq \mathbb{P}(U \leq u \mid S=0)$. The lower bound
is therefore sharp.

For the upper bound, consider $S \sim \mathrm{Bern}(p)$ with $S \perp U$. Then
$\tilde{u} = \mathbb{P}(U \leq u \mid S=1) = \mathbb{P}(U \leq u \mid S=0) = u$, and the upper
bound is also sharp.

\textbf{Step 3: Mean lower bound.}
\begin{align*}
\int_0^1 Q_{Y\mid S=1}\!\!\left(\frac{\max\{u+p-1,0\}}{p}\right) du
&= \int_0^{1-p} Q_{Y\mid S=1}(0)\,du
  + \int_{1-p}^{1} Q_{Y\mid S=1}\!\!\left(\frac{u+p-1}{p}\right) du \\
&= (1-p)\,Q_{Y\mid S=1}(0) + p \int_0^1 Q_{Y\mid S=1}(t)\,dt \\
&= (1-p)\,Q_{Y\mid S=1}(0) + p\,\mathbb{E}(Y\mid S=1). \qquad \square
\end{align*}

\end{document}